\documentclass[lettersize,journal]{IEEEtran}
\usepackage{amsmath,amsfonts}
\usepackage{algorithmic}
\usepackage{algorithm}
\usepackage{array}
\usepackage{subfigure}
\usepackage{hyperref} 
\usepackage{textcomp}
\usepackage{stfloats}
\usepackage{url}
\usepackage{verbatim}
\usepackage{makecell}
\usepackage{multirow}
\usepackage{graphicx}
\usepackage{cite}
\usepackage{amssymb}
\usepackage{color}
\usepackage{xcolor}
\begin{document}

\title{MSM-VC: High-fidelity Source Style Transfer for Non-Parallel Voice Conversion by Multi-scale Style Modeling}

\author{Zhichao~Wang,
    Xinsheng~Wang,
    Qicong~Xie,
    Tao~Li,
    \\Lei~Xie,~\IEEEmembership{Senior Member,~IEEE,} 
    Qiao Tian,
    Yuping Wang
 \thanks{Corresponding author: Lei Xie.}
  
\thanks{Zhichao Wang, Qicong  Xie, Tao Li, and Lei Xie are with the ASLP Lab, School of Computer Science, Northwestern Polytechnical University, Xi’an 710072, China (e-mail: zcwang\_aslp@mail.nwpu.edu.cn, xieqicong@mail.nwpu.edu.cn, taoli@npu-aslp.org, lxie@nwpu.edu.cn).}

\thanks{Xinsheng Wang is with the School of Software Engineering, Xi’an Jiaotong University, Xi’an 710049, China (e-mail: w.xinshawn@gmail.com).}

\thanks{Qiao Tian and Yuping Wang are with the ByteDance SAMI Group, Shanghai 200233, China (e-mail: tianqiao.wave@bytedance.com, wangyuping@bytedance.com).
    }
}

\markboth{Journal of \LaTeX\ Class Files,~Vol.~14, No.~8, August~2021}%
{Shell \MakeLowercase{\textit{et al.}}: A Sample Article Using IEEEtran.cls for IEEE Journals}

\maketitle

\begin{abstract}
In addition to conveying the linguistic content from source speech to converted speech, maintaining the speaking style of source speech also plays an important role in the voice conversion (VC) task, which is essential in many scenarios with highly expressive source speech, such as dubbing and data augmentation. Previous work generally took explicit prosodic features or fixed-length style embedding extracted from source speech to model the speaking style of source speech, which is insufficient to achieve comprehensive style modeling and target speaker timbre preservation. Inspired by the style's multi-scale nature of human speech, a multi-scale style modeling method for the VC task, referred to as MSM-VC, is proposed in this paper. MSM-VC models the speaking style of source speech from different levels, i.e., global, local, and frame levels. To effectively convey the speaking style and meanwhile prevent timbre leakage from source speech to converted speech, each level's style is modeled by specific representation. Specifically, prosodic features, pre-trained ASR model's bottleneck features, and features extracted by a model trained with a self-supervised strategy are adopted to model the frame, local, and global-level styles, respectively. Besides, to balance the performance of source style modeling and target speaker timbre preservation, an explicit constraint module consisting of a pre-trained speech emotion recognition model and a speaker classifier is introduced to MSM-VC. This explicit constraint module also makes it possible to simulate the style transfer inference process during the training to improve the disentanglement ability and alleviate the mismatch between training and inference. Experiments performed on the highly expressive speech corpus demonstrate that MSM-VC is superior to the state-of-the-art VC methods for modeling source speech style while maintaining good speech quality and speaker similarity. Furthermore, ablation analysis indicates the indispensable of every style level's modeling and the effectiveness of each module. 

\end{abstract}


\section{Introduction}

\IEEEPARstart{V}{oice} conversion (VC) aims to modify speech from a source speaker to sound like that of a target speaker while maintaining the linguistic content and speaking style. Traditional VC methods~\cite{GMMToda2007,FWGodoy2012,EBZWU2014} primarily rely on statistical parametric approaches to learn the conversion function between the source and target parallel utterance. Due to the high cost of collecting parallel data, many recent VC approaches~\cite{AutoVCqian2019autovc,AEwang2021adversarially,VAEhsu2016voice,VAEchou2019one,GANVC2018,GANkameoka2018stargan,PPGsun2016phonetic, PPGwang2021one} using non-parallel data have been proposed. Despite recent progress, most voice conversion methods focus on preserving the linguistic content and do not explicitly consider the speaking style of source speaker. In many scenarios, such as dubbing and data augmentation, it is essential to preserve the source speech's speaking style, including emotion, pitch, loudness, and duration. In this paper, we focus on accurately delivering the source speech style in the converted speech while preserving the linguistic content.

One popular approach for style modeling in the VC task is to extract the style embedding of the source speech~\cite{Zhang2020VoiceCascading,liu2020transferring,wang2021enriching,huang2021prosody,du2021identity,du2021expressivevc}, which captures the style information at a global level. Commonly employed strategies to obtain the style embedding include the use of a reference encoder~\cite{ReferenceSkerryRyan2018Reference}, a global style token (GST)~\cite{GSTwang2018style}, and a variational autoencoder (VAE)~\cite{VAEZhang2019VAE}. For instance, in~\cite{liu2020transferring} and~\cite{huang2021prosody}, GST is adopted to learn a high-dimensional representation that encodes source speech style. Du et al.~\cite{du2021expressivevc,distanglemnet} introduce a speech emotion recognition model (SER) trained on an emotion speech corpus to extract a hidden representation to represent source speech style. However, the global level style is too coarse to describe the various aspects of style. Hence, the conversion result may inevitably have a speaking style not so consistent with the source speech. In addition to modeling the global level style information, some efforts also have been conducted from a fine-grained level~\cite{lianZheng2020VCC,Lian2021TowardsFP,gan2022iqdubbing,Speechsplitqian2020unsupervised}. A straightforward way to represent the fine-grained style of source speech is to extract explicit prosodic features~\cite{lianZheng2020VCC}, such as fundamental frequency (f0) and energy. However, handcrafted acoustic features have difficulties to perfectly describe style.
Some studies~\cite{Speechsplitqian2020unsupervised,Lian2021TowardsFP} attempt to model frame-level style representations from the mel spectrogram along with explicit prosodic features, which have demonstrated superiority in the VC task compared to using explicit prosodic features alone. The most recent work~\cite{gan2022iqdubbing} tries to describe the style at the phoneme level by leveraging the transcription of the source speech.

While the above progress has been made in modeling the source speech style, it is insufficient to accurately represent the richness of style information found in human speech at just one or two levels. In general, human speech has a multi-scale nature~\cite{Natureselkirk1986derived} and can be seen as a combination of multi-scale acoustic factors. The style of speech has rich and detailed variations that manifest at different scales. Specifically, we can categorize an utterance based on its speaking style, e.g., reading style, storytelling style, and poetry style, which is based on the style from the global level. From the local level perspective, each speech unit within an utterance, such as syllable or phoneme, has its own characteristics, such as tone, stress, speed, and pause. In addition to the style reflected from the global and local levels, the style can also naturally be reflected at the frame level when speech is represented by frame-level acoustic features. Many previous efforts~\cite{leiyi2021fine,muti-scaleli2021towards,xie2021multi,adaspeechchen2021,wang2021towards,an2019learning} in TTS task have proved the effectiveness of multi-scale style modeling. But most methods primarily focus on modeling style predefined within the corpus. In VC task, modeling arbitrary style without predefined style categories is needed due to the inherent diversity of source speech in practice. Another challenge in VC is the issue of speaker leakage caused by the entanglement of style and speaker timbre~\cite{Zhang2020VoiceCascading,wang2021enriching}, i.e., the speaker's timbre of the source speech is also passed to the converted speech with style representation, consequently impacting the speaker similarity.

With the aim to convey the speaking style of source speech while maintaining the target speaker's identity, this paper proposes a new VC model called MSM-VC. Inspired by the multi-scale nature of human speech, MSM-VC employs a multi-scale style modeling approach that captures style at different levels, i.e., global, local, and frame levels. Considering the unique character of the style reflected from each level and preventing speaker timbre leakage from source speech, each level’s style is modeled by a specific representation. Specifically, the self-supervised learning (SSL) features extracted by vq-wav2vec~\cite{vqwav2vecbaevski2019vq} and bottleneck (BN) features from ASR encoder are used to perform global and local-level style modeling, respectively. As for frame-level style modeling, the prosodic features, including logarithmic domain fundamental frequency (lf0), the short-term average amplitude (energy), and the voice/unvoice flag (VUV), are used. Besides, an explicit constraint module consisting of a speaker classifier~\cite{GANkameoka2018stargan} and a pre-trained speech emotion recognition model (SER)~\cite{liurui2021expressive} is introduced to ensure the retention of the source speech style and target speaker timbre. And inspired by the training process of CycleGAN~\cite{GANkaneko2018cyclegan}, we further employ this explicit constraint module to simulate the style transfer inference process during training to improve the style and speaker disentanglement ability further and alleviate the mismatch between the training and inference process. Experimental results demonstrate that the proposed approach performs superior to the previous state-of-the-art systems on source style modeling while maintaining high speech quality and speaker similarity. Additionally, ablation analysis highlights the importance of each style level, indicating the good design of the proposed model.

Our preliminary work has been presented in \cite{wang2021enriching}, in which only global-level and frame-level style modeling was considered. In this paper, we improved the model's style modeling ability with the proposed multi-scale style modeling module and explicit constraint module. To sum up, the main contributions of this work are as follows:

\begin{itemize}
    \item We propose a novel multi-scale framework for source style modeling in voice conversion. The multi-scale style modeling module is designed to model source speech's style from different levels, i.e., global, local, and frame levels, with a specific feature for each level.
    \item We introduce an explicit constraint model to ensure the retention of the source speech style and target speaker's timbre and meanwhile eliminate the mismatch between training and inference.
\end{itemize}

The rest of this paper is organized as follows. Section~\ref{sc:related work} reviews related work on style modeling. Section~\ref{sc:method} presents the proposed multi-scale source style modeling method for VC. Section \ref{sc:experiments} describes the experimental details. Section \ref{sc:results} presents the experimental results. Section \ref{sc:discussion} discusses the performance and limitations of the proposed method and also the possible future research direction. Finally, Section \ref{sc:conclusion} concludes the paper. Examples of synthesized speech can be found on the project page\footnote{The synthesized samples can be found on \href{https://kerwinchao.github.io/VCStyleModeling.github.io}{\url{https://kerwinchao.github.io/VCStyleModeling.github.io}} \label{demo}}.

\section{Related work}
\label{sc:related work}

\begin{figure*}[htb]	
	\centering
	\includegraphics[width=0.9\linewidth]{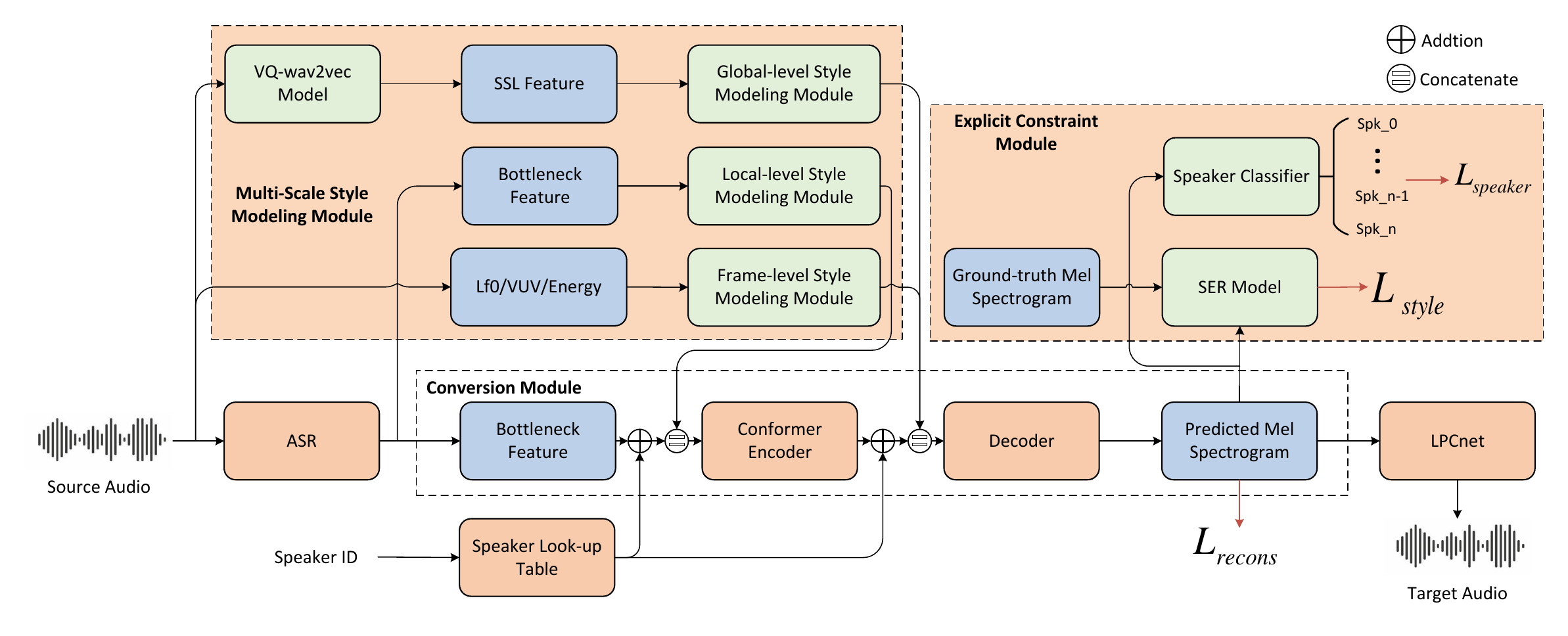}
 \vspace{-3pt}
	\caption{The architecture of the proposed MSM-VC model. Note that local-level style representation is concatenated with BN, and meanwhile, global-level and frame-level style representations are concatenated with the output of the conformer encoder. The explicit constraint model does not involve the inference process. The ASR, vq-wav2vec, SER, and LPCnet are pre-trained models.}
	\label{fig:model_framework}
 \vspace{-7pt}
\end{figure*}

Regarding the different levels of style modeling, this section will review related works on speaking style modeling. Besides, we will also introduce the literature on speaker and style disentanglement.

\vspace{-5pt}
\subsection{Speaking Style Modeling}
Using the style category label to explicitly control the speaking style of synthetic speech is an intuitive way~\cite{EmotioinlabelLee2017EmotionalEN,Emotionlabelluong2017adapting}. But it is limited to predefined style categories in the manually labeled corpus. In contrast, obtaining the style representation from reference speech makes it possible to eliminate the dependence on the explicit label~\cite{ReferenceSkerryRyan2018Reference,GSTwang2018style,VAEZhang2019VAE,du2021expressivevc}. Skerry-Ryan et al.~\cite{ReferenceSkerryRyan2018Reference} propose the reference encoder to extract style embedding with fixed length from reference speech. To make the learned style embedding prominent, global style token (GST)~\cite{GSTwang2018style} model, which extends the reference encoder by adding a style token layer, and variational autoencoder (VAE)~\cite{VAEZhang2019VAE} model are proposed subsequently. Due to the flexibility of these reference speech embedding-based methods, many further efforts are then conducted based on them~\cite{Taoli2021controllable,Zhang2020VoiceCascading,liu2020transferring,appleraitio2020controllable,TPGSTstanton2018predicting,SERcai2021emotion,emotionclassfierwu2019end,huang2021prosody,HGMVAEwei2019,yang2022improvingVAE,emocat}. For instance, the emotion classifier are used to improve the interpretability of style representation learned by GST~\cite{emotionclassfierwu2019end}. And to enhance the style control ability, Raitio et al.~\cite{appleraitio2020controllable} adopt a reference encoder to extract global-level prosodic features, including pitch, energy, and spectral tilt. Some recent work~\cite{ren2020fastspeech,segmentklimkov2019fine,segmentdaxin2020fine,tan2020fine,zhang2021extracting} tries to model fine-grained style representations. The phoneme-level and word-level style representations are two intuitive local-level representations. Fastspeech2~\cite{ren2020fastspeech} adopts a variance predictor to predict the phoneme-level duration, pitch, and energy to represent style in speech. In \cite{zhang2021extracting}, word-level style variation (WSV) is proposed to describe the word-level style, in which WSV is extracted from reference speech during training or text with BERT~\cite{devlin2018bert} during inference. In addition to word and phoneme levels, finer-grained style representations could be obtained from the spectrogram, resulting in frame-level style features~\cite{AutoVCqian2019autovc,du2021identity,lianZheng2020VCC,Lian2021TowardsFP,Speechsplitqian2020unsupervised}. In~\cite{AutoVCqian2019autovc,lianZheng2020VCC}, explicit prosodic features in the frame level are used to represent the style information. Qian et al.~\cite{Speechsplitqian2020unsupervised} and Lian et al.~\cite{Lian2021TowardsFP} utilize an implicit style extractor to extract frame-level style from mel spectrogram to enhance the ability of style modeling.

Compared with the above-mentioned style modeling methods based on a single coarse or fine-grained level, the most recent methods that consider different levels show superiority in style modeling~\cite{leiyi2021fine,muti-scaleli2021towards,xie2021multi,wang2021enriching,adaspeechchen2021,wang2021towards,an2019learning}. For instance, a multi-scale reference encoder is introduced in \cite{muti-scaleli2021towards} to extract the global-level and local-level features from reference speech. In~\cite{leiyi2021fine}, the authors use phoneme-level emotion strength representations and global-level emotion categories to achieve fine-grained emotional speech synthesis. In the multi-speaker and multi-style TTS task~\cite{xie2021multi}, the phoneme-level features, e.g., pitch, duration, and energy, and global style tags are used to model the speaking style. 

Despite the superiority of the multi-scale modeling methods on style modeling, few related efforts of multi-scale style modeling have been conducted on the VC task. Meanwhile, in the absence of ground-truth transcription~\cite{NEURIPS2021_ea159dc9}, accurate pronunciation unit boundaries are unavailable, making modeling local-level style challenging. Besides, most previous methods mainly focus on modeling styles defined in the corpus. However, in practice, the VC system generally has to face arbitrary source speech with a style and speaker that never appears in the training stage, making an effective speaker-style disentanglement method rather than the predefined category-based style modeling method necessary.

\vspace{-1pt}
\subsection{Style and Speaker Disentanglement} 

The speaking style and speaker timbre are highly entangled. It is therefore crucial to squeeze out the source speaker's timbre information while modeling the speaking style. Adversarial training~\cite{Zhang2020VoiceCascading,lee2021styler,wang2021enriching,frameLi2021} is a popular method to squeeze out speaker-related information from style representations, which usually utilizes an auxiliary speaker classifier to predict speaker identity. To suppress the information of the source speaker's identity, this speaker classifier is optimized with adversarial training to make the obtained style embedding speaker-indistinguishable. Besides, constraining the relationship between speaker embedding and style embedding is another popular strategy. For instance, mutual information~\cite{wang2021vqmivc,distanglemnet} and Frobenius norm~\cite{Taoli2021controllable} have been adopted to reduce the correlation between speaker representation and style representation. Qian et al.~\cite{AutoVCqian2019autovc}, Lian et al.~\cite{Lian2021TowardsFP}, and Gan et al.~\cite{gan2022iqdubbing} set the small size bottleneck of style representation to squeeze the speaker information out of the style path. Instead of obtaining the style embedding from the spectrogram, some recent work tries to utilize speaker-irrelevant but style-related features as reference features. For instance, in~\cite{wang2021towards}, the features extracted from a pre-trained ASR model are used to present the speaking style. Lei et al.~\cite{EMP} model speaking style on the perturbed waveform in which the speaker identity has been changed.

Due to the need to face unlabeled and even unseen styles, disentanglement methods based on limited style categories are difficult to apply in our task. Besides, style modeling methods in VC are usually designed in an unsupervised manner without explicit supervision, which makes it difficult to balance the style consistency and speaker similarity of the converted speech. For instance, if the speaking style of the source speech is too different from the speaking style of the target speaker, the speaker similarity will easily be affected. In the training stage, speaking style, linguistic content, and speaker identity all come from the same speech but different speeches during style transfer inference. This mismatch between the training and inference brings insufficient disentanglement and potential performance degradation.

\section{Methodology}
\label{sc:method}

\subsection{Overview}

The proposed MSM-VC is built with a typical encoder-decoder architecture, as shown in Fig.~\ref{fig:model_framework}. This framework consists of three main components: a multi-scale style modeling module, a conversion module, and an explicit constraint module. The multi-scale style modeling module extracts comprehensive style representations from three levels, i.e., global, local, and frame levels. To obtain the global-level and local-level representations, SSL and BN features extracted from source speech are adopted, respectively. Prosodic features, including lf0, VUV, and energy, are used to represent the frame-level style. The conversion module, which consists of a conformer encoder~\cite{gulati2020conformer} and an auto-regressive decoder~\cite{taco2shen2018natural}, takes speaker id, ASR-based content representation BN, and style representations from different levels as input and outputs the mel spectrogram with target speaker timbre and source speaker's speaking style. Besides, to effectively optimize the proposed MSM-VC, an explicit constraint module consisting of a speaker classifier and a pre-trained SER model is introduced in our framework during training. Finally, a modified LPCnet~\cite{valin2019lpcnet} is adopted to reconstruct waveform from mel spectrogram. Note that ASR, vq-wav2vec, SER, and LPCnet are pre-trained models and will not be optimized further during the training of the proposed model.

\begin{figure*}[htb]	
	\centering
	\includegraphics[width=0.9\linewidth]{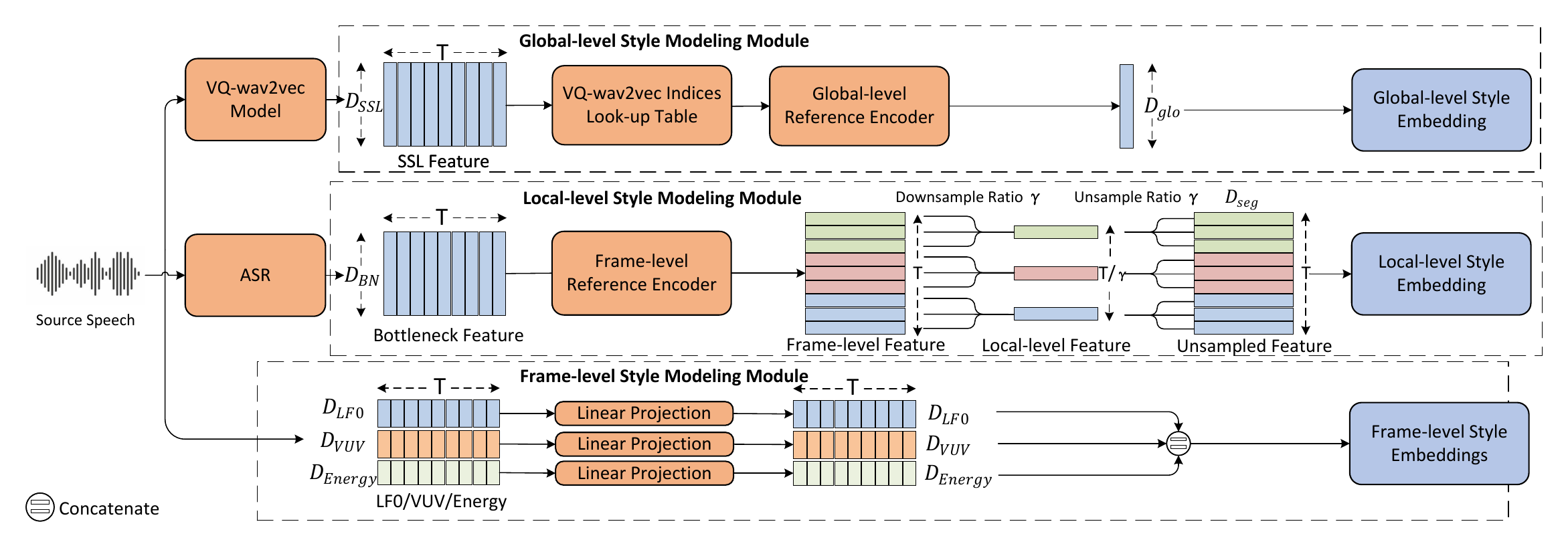}
 \vspace{-5pt}
	\caption{The architecture of the multi-scale style modeling module. Please note that in the figure, we assume $\gamma=3$, we average each group of three vectors and use the averaged vector as the prosodic representation of the current segment. }
	\label{fig:scale_model}
 \vspace{-5pt}
\end{figure*}

\subsection{Global-level Style Modeling}

The global-level style indicates overall speaker style, intensity, and diversity in utterance. The global-level style representation is generally extracted from the mel spectrogram or BN of reference speech using a neural style extractor, e.g., GST, VAE, and reference encoder. However, the mel spectrogram is far from ideal for style modeling due to the redundant information, such as speaker timbre. In contrast, while limited irrelevant acoustic information remained in the BN, the damage to the style information makes it hard to obtain a comprehensive style from BN. Inspired by the characteristic of discrete self-supervised learning feature (SSL) extracted by the self-supervised model vq-wav2vec~\cite{vqwav2vecbaevski2019vq,huang2021anytoonevqwav2vec}, which contains less speaker information than mel spectrogram and richer style information than BN, SSL is adopted for the global-level style modeling. This conclusion will be verified in Section~\ref{sec:richness}.

As shown in Fig.~\ref{fig:scale_model}, the global-level style modeling module consists of three parts, i.e., pre-trained vq-wav2vec model, vq-wav2vec indices look-up table, and global-level reference encoder. The SSL features are first extracted by the pre-trained vq-wav2vec model with the reference speech as input, resulting in features with the dimensionality of $D_{SSL}$ and sequence length of $T$. Here, the SSL feature value indicates the VQ codebook's index. Following Huang et al.~\cite{huang2021anytoonevqwav2vec}, the SSL features are then separated into different groups along the dimension axis to look up different embedding tables, which is helpful for the convergence of training. The resulting embeddings are used as input to the global-level reference encoder~\cite{ReferenceSkerryRyan2018Reference } to obtain the final global-level style embedding. It is essential to note that the SSL feature still contains speaker information which may be conveyed to the converted spectrogram (See Section.~\ref{sec:richness}). In order to prevent the speaker timbre of the source speaker from leaking to the target speech, we set a small bottleneck~\cite{AutoVCqian2019autovc,Lian2021TowardsFP} in the global-level style embedding with the dimension $D_{glo}$. Finally, the obtained global-level style embedding is repeated $T$ times along the time dimension and concatenated with the conformer output.

\subsection{Local-level Style Modeling}
While global-level style information can convey the overall speaking style, the local style expression, e.g., tone, stress, speed, and pause, is also crucial for speaking style. Therefore, it is important to model the local style. Generally, the local style expression is reflected in the speech units, e.g., phonemes or syllables. It is natural to model local style from the phoneme level or syllable level. Unfortunately, the lack of ground-truth transcription in practice makes accurate speech pronunciation unit boundaries inaccessible. To face this challenge, pseudo-speech units are obtained with fixed-length speech segments. While SSL features show superiority in reducing redundant information and maintaining style information, it is not fit to work as the local style modeling feature due to the mispronunciation issue~\cite{van2022comparison}. The discrete process may lead the SSL features of vq-wav2vec to discard some linguistic content. In contrast, the training object of ASR model makes BN extracted from a pre-trained ASR model contain the integrity of the pronunciation information and the consistency within the speech pronunciation unit. Therefore, instead of SSL, here, BN is used for the local-level style modeling.

As shown in Fig.~\ref{fig:scale_model}, we use BN extracted by a pre-trained ASR model as local-level modeling's input with the dimension of $D_{BN}$ and the length of $T$. A modified reference encoder~\cite{Lian2021TowardsFP} is adopted to extract frame-level features. The modified reference encoder consists of six 2D convolutional layers and a GRU layer. The output of the GRU is taken as the frame-level feature, which is then downsampled along the time axis with a fixed ratio $\gamma$ to obtain local-level features. Specifically, taking $\gamma$ frames as a pronunciation unit, we divide the sequence into several segments, and the average of the frames within each segment represents the current pronunciation unit. Then the local-level feature with sequence length of $T/\gamma$ is broadcast-concatenated to the conformer output. As the common duration of consonant-vowel syllables ranges from 150ms to 200ms~\cite{steinschneider2013prosodyrepresentation}, we use $\gamma$ as 16 in practice. To be specific, the duration of each speech pronunciation unit is 200ms with a frameshift of 12.5ms. This pseudo speech unit feature can get rid of the dependency on the transcriptions, making it convenient in the VC task. Meanwhile, the characteristics of BN also ensure that speaker timbre leakage will not happen.

\subsection{Frame-level Style Modeling}

When speech is represented as frame-level acoustic features, e.g., spectrogram, the style naturally varies with the frame. Therefore, in addition to global level and local level, finer grain, i.e., frame-level style, should also be considered. To this end, source speech's explicit acoustic features, including pitch and energy, are adopted. To be specific, lf0 and short-term average amplitude are extracted from source speech to present the pitch and energy, respectively. Besides, VUV, which indicates the frame's voicing, is also used in frame-level style modeling. In practice, lf0 and energy of each utterance are normalized to [0, 1] by utterance-level min-max normalization, which is helpful to prevent naturalness and speaker similarity degradation caused by the unseen style and unseen speaker during the inference process. Normalized energy and lf0 are used to indicate the trend of pitch and energy in the source speech. These features are embedded by linear layers respectively to work as frame-level style embeddings.

\subsection{Explicit Constraint Module}

Achieving high style modeling performance and speaker similarity is an essential goal of the VC task. Since specific representations mentioned above still contain speaker-related information~(See Section.~\ref{sec:richness}), only using them is insufficient to achieve information coupling between speaker and style. Meanwhile, style modeling methods in VC are usually designed in an unsupervised manner without explicit supervision, which makes it difficult to balance the speaking style and timbre, in which the former should be consistent with the source speech while the latter should same as the target speaker. If the speaking style of the source speech is far from that of the target speaker, the speaker similarity will easily be affected. Thus style matching to source speech and speaker similarity to the target speaker should be simultaneously considered to explicitly guide the disentanglement process and balance the source style modeling and target speaker timbre preservation. In this paper, an explicit constraint module consisting of a pre-trained SER model and a speaker classifier is introduced to achieve this end.

\subsubsection{Style matching to source speech}
An intuitive way to constrain the style category is to use a style classification objective function. However, since the speaking styles are distributed in a continuous space, discrete style labels are too coarse to capture finer-grained variation between styles and cannot cover all possible styles. In contrast, representations directly extracted from speech via learnable deep neural networks are considered more suitable as style descriptor~\cite{liurui2021expressive,zhou2021seenESD,SERcai2021emotion}. This representation can capture style-related attributes from a specific utterance, even if the speaking style cannot be accurately represented by manually defined labels. Therefore, the style matching loss here is suitable for measuring the style consistency between style representations of source speech and converted speech. In practice, pre-trained on a style classification task, an SER model is used as a style descriptor to obtain the style representations and calculate the style matching loss.

\subsubsection{Speaker similarity to target speaker}
Since the target speaker's recordings are contained in the training dataset, a speaker classifier is commonly introduced to ensure the speaker similarity to the target speaker, as in Kameoka et al.~\cite{GANkameoka2018stargan}. In practice, the speaker classifier takes the predicted mel spectrogram as input and outputs the probability of this spectrogram belonging to the target speaker identity.

Generally, in the training phase, all conditional information, including speaking style, content, and speaker identity, comes from the same utterance and is used to reconstruct the original utterance. However, in the style transfer scenario, i.e., the inference stage, the content and speaking style are from the source speech, while the speaker identity is from the target speaker, resulting in inconsistency between these two phases. This inconsistency could make it hard to measure the disentanglement ability during training and result in limited performance for the VC task. To solve this problem, inspired by the training process of CycleGAN~\cite{GANkaneko2018cyclegan}, applying the explicit constraint module allows us to divide the training process into two different modes, i.e., \textit{reconstruction mode}, and \textit{simulation mode}. In the reconstruction mode, the whole model is trained using paired data with style, content, and speaker id of the same utterance. In contrast, to perform style transfer simulation, style and content are extracted from the same utterance, while the speaker id is randomly assigned. Without ground-truth utterance in simulation mode, the explicit constraint module plays a core role in the balance of style modeling and speaker timbre preservation.

\subsection{Objective Functions}
\label{sec:objective}

To effectively optimize the proposed model, style matching loss, speaker classification loss, and mel reconstruction loss are introduced to ensure style consistency with the source speech, speaker timbre similarity with the target speaker, and the reconstruction quality of the mel spectrogram, respectively.

\subsubsection{Style matching loss}

\begin{figure}[htb]
	\centering
	\includegraphics[width=1.0\linewidth]{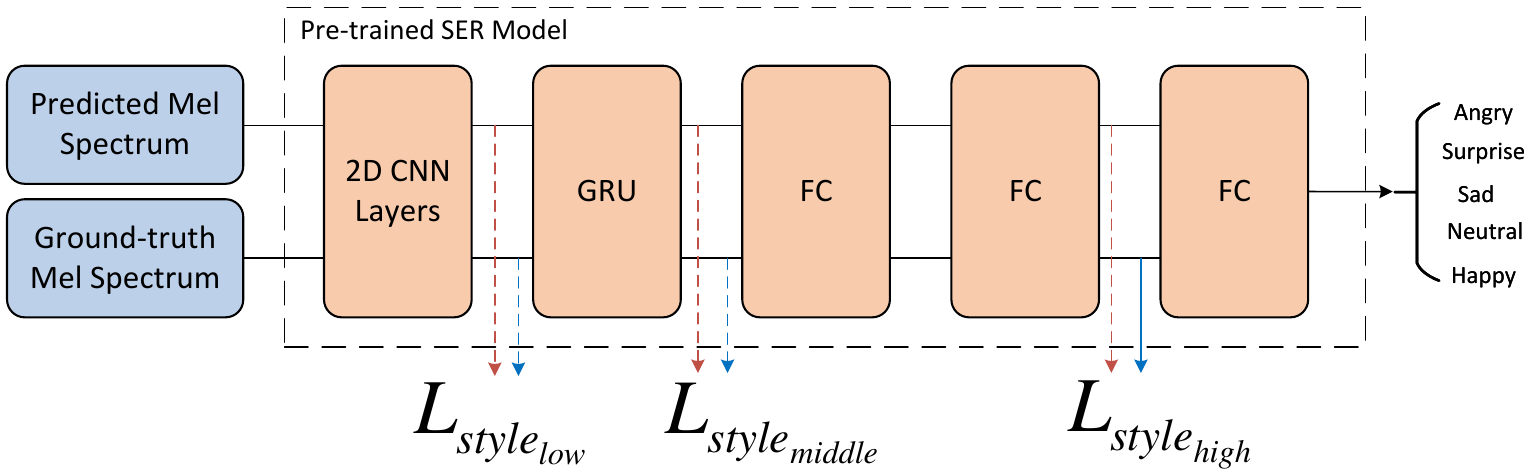}
	\caption{The network architecture of the SER model. The features of the three network layers are extracted to calculate the style matching loss.}
	\label{fig:ser_model}
\end{figure}

The matching loss measured by the SER model ensures that the converted speech has the same style as the source speech. As shown in Fig.~\ref{fig:ser_model}, to constrain the style from different levels, style-related features extracted from different layers of the SER model are obtained. Following~\cite{liurui2021expressive}, 2D convolution extracts a variable-length hidden representation $h_{low}$ from the mel spectrogram $Y$, and the GRU extracts the fixed-length vector $h_{middle}$ from the temporal information of $h_{low}$. Besides, we take the hidden representation calculated by the second FC layer as $h_{high}$. \textit{Low}, \textit{middle} and \textit{high} stand for the abstraction degree of the hidden representation of the style. Finally, with the features extracted from different layers, the style matching loss between the source mel spectrogram $Y$ and the predicted mel spectrogram $\hat{Y}$ can be defined as
\begin{equation}
    \mathcal{L}_{style}=\sum_{s}^{}||h_s-\hat{h_s}||^{2}_{2},s\in \{ low, middle, high \}
\end{equation}
where $h$ and $\hat{h}$ are extracted from the SER model using $Y$ and $\hat{Y}$ as input, respectively.

\subsubsection{Speaker classification loss}
The architecture of the speaker classifier is the same as the SER model. Same as the practice in \cite{GANkameoka2018stargan}, it takes the predicted mel spectrogram as input to predict the current speaker identity. The corresponding speaker classification loss is defined as:
\begin{equation}
    \mathcal{L}_{speaker} = -\log{P(s|\hat{Y})} 
\end{equation}
where $s$ represents the given target speaker label, $\hat{Y}$ is the predicted mel spectrogram, and $P(s|\hat{Y})$ represents the probability of being identified as speaker $s$ under the condition of inputting $\hat{Y}$. 

\subsubsection{Mel reconstruction loss}

The mel reconstruction loss, working as a basic objective function for the speech synthesis, is to make the model create reasonable target speech based on style, speaker, and content. L2 distance between predicted mel spectrogram $Y$ and ground-truth mel spectrogram $\hat{Y}$ is adopted as the mel reconstruction loss, which is defined as:
\begin{equation}
  \mathcal{L}_{recons} = ||Y-\hat{Y}||^{2}_{2}
\end{equation}

\subsubsection{Overall objective function}
The overall objective function is described as follows:
\begin{equation}
\begin{aligned}
\mathcal{L}_{total}=\alpha*&\mathcal{L}_{recons}+\mathcal{L}_{speaker}+\alpha*\mathcal{L}_{style_{low}}\\
&+\mathcal{L}_{style_{middle}}+\mathcal{L}_{style_{high}}
\end{aligned}
\end{equation}
where the value of $\alpha$ is $0$ or $1$ to indicate the simulation mode and the reconstruction mode of model training, respectively. In particular, due to the lack of ground-truth mel spectrogram in the simulation mode, the model is only optimized with $\mathcal{L}_{style}$ and $\mathcal{L}_{speaker}$ due to the lack of ground-truth mel spectrogram. Note that, due to the high similarity between $h_{low}$ and mel spectrogram~\cite{liurui2021expressive}, the low-level style feature $h_{low}$ contains rich speaker-related information. Therefore, the $h_{low}$-based style loss $\mathcal{L}_{style_{low}}$ is neither considered in the simulation mode to avoid the effect on the speaker similarity.

In practice, we first train the model in reconstruction mode ($\alpha=1$) until the model is converged. Then, we finetune the trained model in both reconstruction and simulation modes. Since only training exists in the simulation mode, where no ground-truth mel spectrogram is available for mel reconstruction loss, the model attends to fit the other training objectives and ignores the mel reconstruction ability which has been learned in the first stage. Thus, following the similar process of CycleGAN~\cite{GANkaneko2018cyclegan}, to ensure the model's mel reconstruction ability, the reconstruction mode is introduced in the finetune stage and used alternately with the simulation mode. Besides, only the decoder is updated in this stage to further prevent the forgetting of learned reconstruction ability, since intuitively more training parameters are more likely to lead to overfitting~\cite{boffintts,gctts}.

\section{Experimental Setup}
\label{sc:experiments}

To evaluate the performance of MSM-VC on the VC task, experiments are conducted on a Chinese multi-speaker speech corpus. In this section, the databases for the voice conversion model and also for the pre-trained models will be introduced. Besides, implementation details, compared methods, and the evaluation method will also be introduced.

\subsection{Corpus}

An internal multi-speaker speech corpus licensed from Databaker\footnote{https://www.data-baker.com/data/index/compose} is adopted to evaluate the proposed method. This corpus is a standard Mandarin reading corpus recorded by 57 professional voice actors, including 30 females and 27 males. Each speaker performs 500 utterances, resulting in a total duration of 42h. One female speaker labeled with $s1$ in this corpus is used as the target speaker. A test set contains a series of highly expressive speech in different scenarios, including emotions, movies, novels, daily conversation, and variety shows. Twenty sentences are randomly selected from 6 kinds of emotions and the other four speaking styles, respectively, resulting in 200 utterances for the test. In the experiments, the ASR model is trained with 10k hours of speech from Wenetspeech~\cite{zhang2022wenetspeech}. The SER model is trained with the open-source emotional data ESD~\cite{zhou2021seenESD}, which is recorded by 20 Chinese and English speakers by performing five kinds of emotions, i.e., Angry, Happy, Neutral, Sad, and Surprise. The vq-wav2vec is an open-source pre-trained model trained on the 960h Librispeech dataset~\cite{panayotov2015librispeech}.

\subsection{Implement Details}

All speech utterances are downsampled to 16kHz and represented by 80-dim mel spectrogram which is computed with 50ms frame length and 12.5ms frame shift. The ASR system is a TDNN-F model implemented by Kaidi toolkits~\cite{povey2011kaldi}. We use the 256-dim bottleneck features ($D_{BN}=256$) as the linguistic representation, which is extracted from the last fully-connected layer before softmax. The officially released vq-wav2vec model\footnote{https://github.com/pytorch/fairseq} is used to extract 2-dim SSL features ($D_{SSL}=2$). Pyworld toolkit\footnote{https://github.com/JeremyCCHsu/Python-Wrapper-for-World-Vocoder} is adopted to extract F0. Note that lf0, VUV, and energy used in this paper are all 1-dim features ($D_{LF0}=D_{VUV}=D_{Energy}=1$). Modified LPCnet\cite{valin2019lpcnet} based on official implementation\footnote{https://github.com/mozilla/LPCNet} is adopted to reconstruct waveform from mel spectrogram. We use ground-truth mel spectrogram to train the modified LPCnet on the multi-speaker corpus and finetune it with data of the target speaker $s_{1}$.

The conformer encoder consists of one conformer block, which contains eight heads of multi-head attention module, convolution module with 31 kernel size, a feed-forward module with one expansion factor and the settings of other parts remain the same as the official setting\footnote{https://github.com/sooftware/conformer}. The architecture and hyperparameters of the global-reference encoder keep the origin configuration\cite{ReferenceSkerryRyan2018Reference}. To be specific, it consists of 6 convolution layers and a GRU layer. Each convolution layer is composed of 3×3 filters with 2×2 stride, SAME padding, and ReLU activation. The number of filters in each layer is 32, 32, 64, 64, 128, and 128, respectively. Batch normalization is applied to each layer. The output of convolution layers is fed into the GRU with four units ($D_{glo} = 4$). Different from the global-level reference encoder, the frame-level reference encoder adopts convolution layers composed of $3\times3$ filters with $1\times2$ stride and the outputs of the GRU at every timestep form the frame-level representation ($D_{seg}=4$). The decoder is an auto-regressive module~\cite{taco2shen2018natural} which consists of prenet, decoder RNN, and postnet. In the reconstruction training stage, the conversion model is trained for 240 epochs with batch size of 32. Adam optimizer is used to optimize the model with learning rate decay, which starts from $1\times10^{-3}$ and decays every 20 epochs with decay rate of 0.7. In the simulation training stage, the conversion model is trained for 70 epochs, in which process the learning rate starts from $1\times10^{-6}$ and decays every 20 epochs with decay rate of 0.5.

\subsection{Compared Methods}
To evaluate the performance of the proposed method \textbf{MSM-VC} on the VC task, three recent state-of-the-art systems designed for VC are compared in the experiments. These compared methods represent three typical VC approaches, i.e., global-level reference embedding based method, frame-level representation based method, and a hybrid strategy. Note that all systems use the same vocoder LPCNET to reconstruct waveform from the mel spectrogram. Details of these compared methods are introduced as followings.

\textbf{GST-VC}~\cite{liu2020transferring} is a typical global reference representation-based VC method. In this model, GST~\cite{GSTwang2018style} extracts global style information from the source speech's mel spectrogram. As for the linguistic content of source speech, a pre-trained ASR model is used to obtain the phoneme sequence. Then, the converted speech is produced conditioned on the phoneme sequence, global style information, and target speaker identity information. 

\textbf{REF-VC}~\cite{Lian2021TowardsFP} is a frame-level style representation-based method. In this model, the authors utilize a modified reference encoder to learn frame-level style representation from the mel spectrogram in an unsupervised manner. The speaking style of source speech is conveyed to the converted result by this learned frame-level style representation together with f0.

\textbf{Hybrid-VC}~\cite{wang2021enriching} is the preliminary work of the current method, in which explicit prosodic features (lf0 and energy) together with global-level style representation are used to model the speaking style. Unlike the current work, this preliminary method simply describes style from two levels and lacks explicit supervision for style and speaker.

\subsection{Evaluation Metrics}
With the input of source speech, the goal of the VC task is to obtain the converted speech that shares the same linguistic content and speaking style as the source speech but with the timbre of the target speaker. Therefore, there are two aspects that should be considered in the evaluation: 1) the style similarity between the source speech and converted speech; 2) the speaker similarity between the target speaker and that of converted speech. Besides, as a kind of speech synthesis task, the quality of the produced speech should also be evaluated. To evaluate the converted speech from the above three aspects, both objective and subjective evaluation methods are conducted in the experiment.

\subsubsection{Objective metrics}

Considering both lf0 and energy are style-related acoustic features, Pearson correlation coefficients of lf0 and energy are calculated between the source speech and convert speech to objectively reflect the style similarity. Higher Pearson correlation coefficients of lf0 or energy indicate better style similarity. As for the evaluation of speaker similarity, a pre-trained speaker verification system~\cite{ECAPA_TDNN} trained on CN-Celeb~\cite{cnceleb} is introduced. Cosine similarity between the SV model-based speaker embeddings of converted speech and speech from the target speaker shows speaker timbre similarity between them. Higher cosine similarity means better similarity between speaker timbres of converted speech and the target speaker.

\subsubsection{Subjective metrics} 

In addition to the objective evaluation, a human perceptual rating experiment is performed to evaluate the converted speech in terms of style similarity (between converted speech and source speech), speech quality, and speaker similarity (between converted speech and target speaker speech). To facilitate the comparison between different models in terms of style modeling, a comparative mean opinion score (CMOS) test is also performed in the experiment.
In the test, given the reference speech, listeners are asked to rate whether the first sample is better or worse than the second one in terms of style similarity, using a seven-point scale comprised of +3 (much better), +2 (better), +1 (slightly better), 0 (same), -1 (slightly worse), -2 (worse), -3 (much worse). Besides, A/B preference is also presented, which can be simultaneously obtained during the CMOS test, to provide another perspective into the comparison. Different from the CMOS test, participants are asked to choose the better one or ``Neutral" when both samples are similar in A/B preference. As for the evaluation of speaker similarity and speech quality, following the typical mean opinion score (MOS) test method, listeners are asked to rate a given speech a score ranging from one to five for its speaker similarity or speech quality. A higher score means better performance, and score value 1 means very bad, and 5 means excellent. In the experiments, 60 utterances are randomly selected from the test set, and 20 participants in total join in both CMOS and MOS tests. 

\section{Experimental Results}
\label{sc:results}

Experimental results, including the comparison with other methods and ablation studies, will be presented in this section. Besides, the rationality of the feature choosing for different-level style modeling is also presented. We also investigate the model's behavior in two training modes and the model size of different systems. We highly recommend readers listen to the converted samples from \url{https://kerwinchao.github.io/VCStyleModeling.github.io}.

\subsection{Subjective Evaluations}

\begin{table}[]
\centering
\caption{CMOS and A/B preference results for the comparison of the proposed method with other methods in terms of style modeling. A positive CMOS value means that the proposed method is better than the compared method and vice versa. $p$ denotes p-value to verify the significance of the results.}
\label{tab:CMOS}
\setlength{\tabcolsep}{1.5mm}
\renewcommand\arraystretch{1.5}
\centering
\begin{tabular}{c|c|ccc|c}
\hline
                   \multirow{2}{*}{}    & \multirow{2}{*}{\makecell[c]{Style CMOS}}    & \multicolumn{4}{c}{ \makecell[c]{Preference (\%)}}       \\ \cline{3-6}
            &      & \makebox[0.06\textwidth][c]{\makecell[c]{Compared\\ Method}}      & \makebox[0.05\textwidth][c]{ Neutral}    & \makebox[0.05\textwidth][c]{MSM-VC}  &$p$                                                    \\ \hline                   
GST-VC & 1.393    & 5.9      &11.6 &82.5  &\textless 0.01          \\ 
REF-VC                & 0.507                &20.0 &26.2   &53.8 &\textless 0.01       \\ 
Hybrid-VC         & 0.461                &21.6 &28.4   &50.0 &  \textless 0.01     \\ \hline
\end{tabular}
\vspace{-10pt}
\end{table}

\subsubsection{Source style modeling performance}
Table~\ref{tab:CMOS} presents the comparison of the style modeling performances between the proposed and compared methods, in which subjective evaluation with CMOS is reported. In this subjective rating test, participants have to rate two compared samples, in which one is obtained by the proposed method and another one is from a compared method. A positive value means that the proposed method is better than the compared one and vice versa. Besides, the A/B preference test is also presented to give further evaluation between the compared methods and the proposed method.

As shown in this table, all CMOS values are larger than 0, which means that compared with all of these listed methods, the proposed method shows superiority in style modeling. And the scores of A/B preference also show that MSM-VC significantly outperforms the compared methods (\textit{p-value} smaller than 0.01). These results demonstrate the effectiveness of the proposed method on style preservation. When we pay attention to the specific CMOS and preference scores compared with different methods, it can be found that the largest performance gap exists between the proposed method and GST-VC, indicating the inferiority of GST-VC in style modeling. This poor performance demonstrates that only modeling the style from a global coarse-grained is insufficient for style preserving in the VC task. In contrast, modeling the style from a fine-grained level, e.g., the frame-level-based method REF-VC, results in better performance. However, no matter the coarse-grained style modeling method or fine-grained style modeling method, this single-level modeling method is inferior to Hybrid-VC and the proposed method, which indicates the importance of modeling style from different levels.

\subsubsection{Speech quality and speaker similarity}

In addition to style modeling ability, speech quality and speaker similarity are also important aspects to evaluate the performance of a VC model. The results of MOS tests in terms of speech quality and speaker similarity for different models are shown in Table~\ref{tab:mos}. Compared with REF-VC and Hybrid-VC, which are obviously superior to GST-VC in terms of style modeling, the proposed method achieves better MOS scores both in speech quality and speaker similarity. Moreover, significant tests confirm that MSM-VC outperforms REF-VC ($p$-value smaller than 0.05) but no significant difference is shown between the MOS results of MSM-VC and Hybrid-VC.

Compared with the proposed method, GST-VC gets better MOS values in speech quality and speaker similarity. This good performance is attributable to the phoneme-based content modeling method and global-level modeling strategy. To be specific, using the phoneme sequence as the content input can filter out all information that is unrelated to the content from the source speech, thus preventing the effect from the source speech to the final results. Besides, the global-level style embedding only indicates an overall style, which could bring very limited noise information from the source speech to the converted speech. However, this good performance in the speech quality and speaker similarity is at the expense of style modeling because of the limited information taken from the source speech. Furthermore, the significance test result between GST-VC and MSM-VC shows that the differences are not significant ($p$-value greater than 0.05) in the speech quality and speaker similarity. It demonstrates that MSM-VC can maintain high speech quality and speaker similarity while achieving modeling source style.

\begin{table}[]
\caption{Comparison of the proposed method with GST-VC, REF-VC, and Hybrid-VC in terms of speech quality and speaker similarity MOS with confidence intervals of 95$\%$. The bold indicates the best performance out of the four models. $p$ denotes the p-value between the comparison and proposed systems.}
\centering
\renewcommand\arraystretch{1.5}
\begin{tabular}{c|cccc}
\hline
\centering
\label{tab:mos}
\multirow{2}{*}{} & \multicolumn{4}{c}{MOS$~(\uparrow)$}                                  \\ \cline{2-5} 
                  & Speech Quality & \multicolumn{1}{c|}{$p$} & Speaker Similarity & $p$ \\ \hline
GST-VC            & \textbf{3.60$\pm$0.094} & \multicolumn{1}{c|}{0.331}    & \textbf{3.71$\pm$0.106}  & 0.141       \\
REF-VC            & 3.40$\pm$0.106   & \multicolumn{1}{c|}{0.014}   & 3.46$\pm$0.080    & 0.011     \\
Hybrid-VC         & 3.53$\pm$0.117  & \multicolumn{1}{c|}{0.579}   & 3.60$\pm$0.194     & 0.470    \\
MSM-VC            & 3.54$\pm$0.083  & \multicolumn{1}{c|}{-}  & 3.66$\pm$0.086  & -       \\ \hline
\end{tabular}
\vspace{-10pt}
\end{table}

\subsection{Objective Evaluations}

\begin{table}[]
\centering
\label{tab:objective}
\caption{Objective comparison of different models on the VC task in terms of style modeling and speaker similarity. Note that $0.881$ are calculated from the target speaker data.}
\setlength{\tabcolsep}{2mm}
\renewcommand\arraystretch{1.5}
\begin{tabular}{c|cc|c}
\hline
\multirow{2}{*}{} & \multicolumn{2}{c|}{Pearson Coefficient~$(\uparrow)$}  & \multirow{2}{*}{\makecell[c]{Cosine Similarity\\$(\uparrow, 0.881)$}} \\ \cline{2-3}
                  & \multicolumn{1}{c|}{\makebox[0.06\textwidth][c]{Lf0}}      & Energy                               &                                   \\ \hline
GST-VC            & \multicolumn{1}{c|}{0.633}    & 0.820                             & \textbf{0.828}                             \\
REF-VC            & \multicolumn{1}{c|}{0.738}    & 0.947                             & 0.791                             \\
Hybrid-VC         & \multicolumn{1}{c|}{0.742}    & \textbf{0.971}                             & 0.811                             \\
MSM-VC            & \multicolumn{1}{c|}{\textbf{0.757}}    & 0.968                              & 0.823                             \\ \hline
\end{tabular}
\end{table}

The objective comparison among different models is shown in Table~\ref{tab:objective}, in which a higher Pearson correlation coefficient indicates the speaking style of source speech is better reflected in the converted speech. As can be seen from this table, while the energy Pearson coefficient of MSM-VC is slightly lower than Hybrid-VC, the proposed system achieves overall better scores than other compared methods, which is consistent with the subjective evaluation results. 

The objective comparison of different models on the speaker similarity also presents similar results to that obtained in the subjective evaluation test. To be specific, GST-VC gets the highest speaker similarity, which means that GST-VC has the best performance in achieving converted speech with the target speaker's timbre. The proposed MSM-VC ranks next to GST-VC and shows better performance than REF-VC and Hybrid-VC. As discussed in the subjective evaluation and also the worst performance of GST-VC in the subjective style modeling evaluation, the proposed method presents the most balanced performance in the style modeling and speaker similarity, demonstrating the superiority of MSM-VC in source style modeling while achieving high speaker similarity.

\subsection{Component Analysis}

In this section, ablation studies will be conducted to validate the effectiveness of each component of MSM-VC, i.e., the multi-scale style modeling module, explicit constraint module, and the simulation strategy for the training of MSM-VC.

\subsubsection{Effectiveness of different style level}

\begin{table}[]
\centering
\caption{CMOS and A/B preference results for evaluating the effect of different style levels on the source style modeling. $p$ denotes p-value to verify the significance of the results.}
\label{tab:ablation_cmos}
\setlength{\tabcolsep}{1.5mm}
\renewcommand\arraystretch{1.5}
\centering
\begin{tabular}{c|c|ccc|c}
\hline
                   \multirow{2}{*}{}    & \multirow{2}{*}{\makecell[c]{Style CMOS}}    & \multicolumn{4}{c}{ \makecell[c]{Preference (\%)}}       \\ \cline{3-6}
            &      & \makebox[0.06\textwidth][c]{\makecell[c]{Compared\\ Method}}      & \makebox[0.05\textwidth][c]{ Neutral}    & \makebox[0.05\textwidth][c]{MSM-VC}  & $p$                                                  \\ \hline                   
w/o Global & 0.40    & 17.3      &37.4 &45.3  &\textless 0.01          \\  
w/o Local        & 0.44                &17.4 &32.0   &50.6 & \textless 0.01       \\ 
w/o Frame         & 1.06                 &9.4 &12.0   &78.6 & \textless 0.01     \\ \hline
\end{tabular}
\vspace{-10pt}
\end{table}

As we argue that human speech's multi-scale nature makes it is necessary to model the speaking style from different levels, in this section, we would like to analyze the effectiveness of style modeling from each level on the VC task. To be specific, several variants of MSM-VC are evaluated by dropping one of the style modeling levels. As shown in Table~\ref{tab:ablation_cmos}, \textit{w/o Global}, \textit{w/o Local}, and \textit{w/o Frame} indicate a variant of MSM-VC without modeling the global-level style, local-level style, and frame-level style, respectively. In this table, the performances of those MSM-VC variants are compared with MSM-VC using the CMOS test, in which a positive CMOS score means MSM-VC is better than the compared variant. A/B preference and significant tests also are conducted in this comparison.

As can be seen from this table, all CMOS values are positive, indicating that without modeling the style from any level will decrease the performance of style modeling. And the scores of A/B preference also show similar results. Among the three levels for style modeling, the frame-level modeling module shows the most important role in style modeling, by dropping which the CMOS value is larger than 1. This obvious effect is attributed to those frame-level style representations, i.e., lf0, VUV, and energy, which are able to represent fine-grained style from different perspectives directly. The global-level modeling module and local-level modeling module also play important roles in the final style modeling, which can be demonstrated by the large CMOS values that are larger than 0.4. All these results show the necessity to model the speaking style from different levels and also indicate the good design of the proposed multi-scale modeling method.

\subsubsection{Effectiveness of explicit constraint module}

\begin{table}[]
\centering
\caption{Ablation analysis of explicit constraint module and simulation training method}
\label{tab:explicit}
\setlength{\tabcolsep}{1.5mm}
\renewcommand\arraystretch{1.5}
\centering
\begin{tabular}{c|c|c|c|c}
\hline
                   \multirow{2}{*}{}    & \multirow{2}{*}{\makecell[c]{Cosine\\Similarity\\$(\uparrow, 0.881)$}}    & \multicolumn{2}{c|}{ \makecell[c]{Pearson Coefficient~$(\uparrow)$}}  &  \multirow{2}{*}{\makecell[c]{Speech\\Quality$(\uparrow)$}} \\ \cline{3-4}
            &      & \makebox[0.06\textwidth][c]{Lf0}      & \makebox[0.05\textwidth][c]{Energy}   &                                                   \\ \hline                   
\makecell[c]{w/o Speaker \\ Classifier} & 0.794   &0.748  & 0.908          &  3.56$\pm$0.081       \\ 
w/o SER                & 0.827                &0.715 &0.883    & 3.58$\pm$0.089        \\ 
w/o Simulation         & 0.810                &0.724 &0.906   & 3.56$\pm$0.072        \\ 
MSM-VC                 & 0.823                &0.757 &0.968   &3.54$\pm$0.083         \\\hline
\end{tabular}
\vspace{-10pt}
\end{table}

An ablation study to analyze the effectiveness of the explicit constraint module and simulation training stage driven by explicit constraints is also conducted. The results are shown in Table~\ref{tab:explicit}, in which the performances of two MSM-VC variants obtained by dropping speaker classifier and SER, referred to as \textit{w/o Speaker Classifier} and \textit{w/o SER} respectively, are presented. Besides, the performance of MSM-VC trained without the simulation stage, named \textit{w/o Simulation}, is also compared. 

As shown in this table, the speaker classifier and SER show obvious effects on speaker similarity and style consistency, respectively. To be specific, the dropping of the speaker classifier brings a 3.5\% relative decrease compared with MSM-VC in speaker cosine similarity. After removing the SER, the pearson coefficients of lf0 and energy decrease by 5.5\% and 6.2\% compared with the proposed method. As for the speech quality, a tiny negative impact exists when the speaker classifier or SER module is utilized. However, this negative impact is very limited, indicating the effectiveness of the speaker classifier and SER in modeling the speaker timbre and style while maintaining the quality of synthesized speech.
Furthermore, as shown in Table~\ref{tab:explicit}, the simulation training stage also shows a positive effect on speaker similarity and style modeling. The model trained without the simulation stage presents a 1.5\% speaker similarity drop. The pearson coefficients of lf0 and energy also show 4.3\% and 6.4\% relative decrease. While a slight performance decrease appears when the simulation training stage is adopted in terms of speech quality, similar to the effect of using speaker classifier and SER, this effect is quite tiny. The reason behind this speech quality decrease caused by the explicit constraint is straightforward, which is caused by the lack of reconstruction constraints for performing these explicit losses. However, the obvious improvements to the speaker modeling and style modeling but a slight decrease in the speech quality demonstrates the effectiveness of the explicit constraint module in the VC task.

\subsection{Analysis of Feature Choosing for Style Modeling}
\label{sec:richness}

\begin{table}[]
\centering
\caption{Comparison of different features in terms of style and speaker richness.}
\label{tab:richness}
\setlength{\tabcolsep}{3.5mm}
\renewcommand\arraystretch{1.5}
\begin{tabular}{c|cc|c}
\hline
\multirow{2}{*}{} & \multicolumn{2}{c|}{MSE~$(\downarrow)$}           & \multirow{2}{*}{Speaker Accuracy~$(\uparrow)$} \\ \cline{2-3}
                  & \multicolumn{1}{c|}{Lf0}  & Energy &                                   \\ \hline
Mel               & \multicolumn{1}{c|}{1.25} & 0.032  & 0.93                              \\
BN                & \multicolumn{1}{c|}{2.14} & 0.078  & 0.48                              \\
SSL               & \multicolumn{1}{c|}{1.70} & 0.041  & 0.73                              \\
SSL~(EN)               & \multicolumn{1}{c|}{1.64} & 0.045  & -                              \\\hline
\end{tabular}
\end{table}

Using the appropriate feature for style modeling of a specific level is important. In MSM-VC, SSL and BN features are adopted for the global and local-level style modeling. Here, we would like to show the information richness of these features in terms of style and speaker information. In addition to the BN and SSL features, mel spectrogram, which is the commonly used acoustic feature, is also compared. 
Due to the lack of direct indicators of information richness, these features are compared in several prediction tasks, alternatively. Specifically, each kind of feature is used to train models to predict f0, energy, and speaker, respectively. The predicted accuracy of f0 and energy are evaluated by the MSE between the predicted results and ground truth. A lower MSE of f0 and energy means that more style information is contained in the feature. The speaker classification performance is evaluated by the classification accuracy, and higher speaker accuracy indicates more contained speaker information. We use the multi-speaker corpus mentioned in Section~\ref{sc:experiments} for the model training and randomly select 50 sentences from each speaker as the test set. The model's structure is the same as the SER.

As shown in Table~\ref{tab:richness}, the mel spectrogram achieves the best performance in speaker classification and prosodic feature prediction, which is as expected due to the least acoustic information loss compared with BN and SSL features. This high speaker-related correlation makes mel spectrogram a non-ideal styling modeling feature because of the speaker leakage issue caused by the speaker information. In contrast, BN and SSL show less speaker information. The characteristic of SSL has richer style and speaker information than BN but less speaker information than mel spectrogram making SSL suitable for global style modeling, in which the obtaining of the global style embedding could effectively reduce the speaker-related information. As for the modeling of local-level style modeling, more information could be kept in the final style embedding, thus making the feature with rich speaker information unsuitable. Therefore, the BN feature, which contains the least speaker information, can prevent the speaker leakage issue when it is taken for the local-level style modeling. While less style information is contained in BN and SSL compared with mel spectrogram, modeling the style from different levels can effectively bridge this gap. Furthermore, considering that the vq-wav2vec model is trained on English data, we also tested the style richness of SSL extracted from English data, referred to as \textit{SSL(EN)}. Fifty-seven speakers from VCTK~\cite{VCTK} are selected for training and 50 sentences from each speaker are used as the test set. As shown in Table VI, SSL(EN) extracted from English speech has a similar style richness to that extracted from Chinese speech, which indicates the language independence of style information carried by the SSL feature.

\begin{figure*}[ht]
	\centering
	\footnotesize
	\centering
		\begin{minipage}{0.35\linewidth}
		    \subfigure[Mel Reconstruction~($L_{recons}$)]{
			\includegraphics[width=\textwidth]{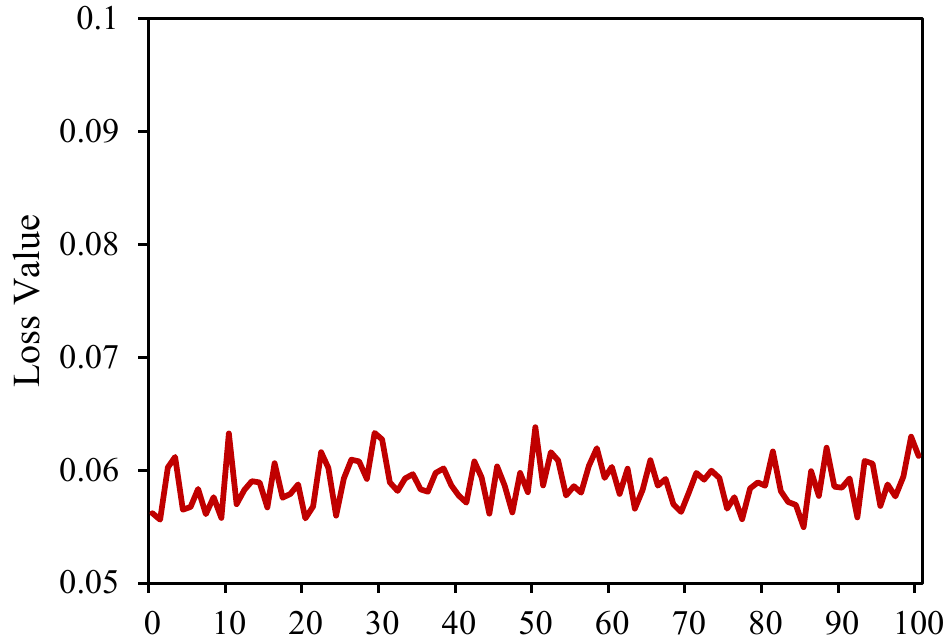}}
		\end{minipage}
		\vspace{0.1cm}
		\hfill
		\centering
		\begin{minipage}{0.31\linewidth}
		\subfigure[Style Matching~($L_{style}$)]{
			\includegraphics[width=\textwidth]{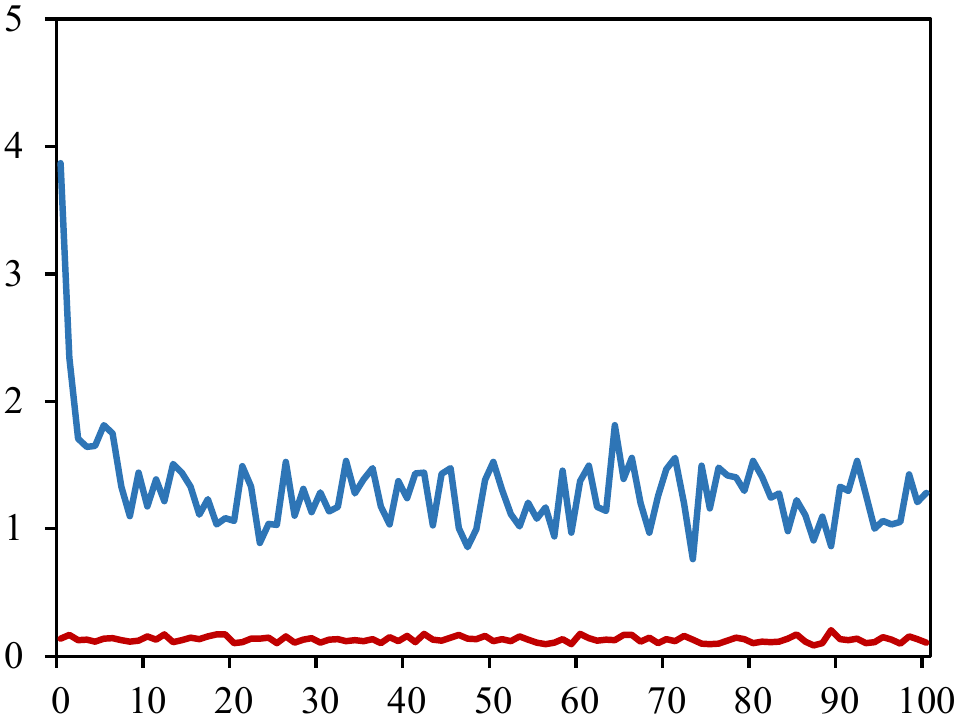}}
		\end{minipage}
		\centering
		\begin{minipage}{0.31\linewidth}
		\subfigure[Speaker Classification~($L_{speaker}$)]{
			\includegraphics[width=\textwidth]{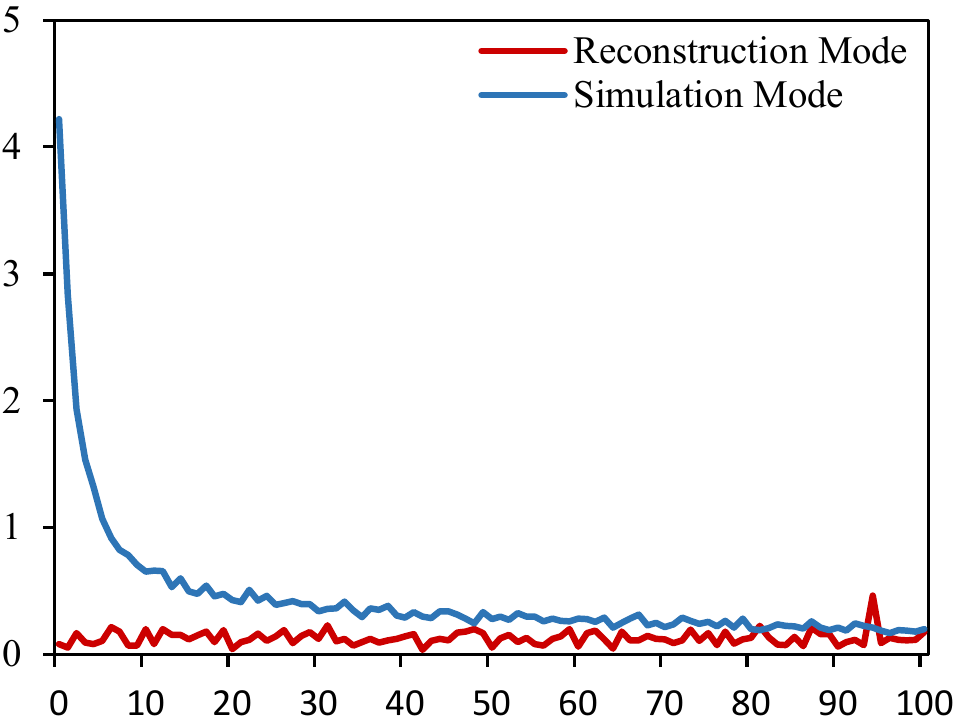}}
		\end{minipage}
		\vfill
	\caption{Visualization of loss curves for reconstruction and simulation modes in finetune stage.}
 \vspace{-5pt}
	\label{figloss}
\end{figure*}

\vspace{-5pt}
\section{discussion}
\label{sc:discussion}

In this paper, a multi-scale style modeling method for the VC task, named MSM-VC, is proposed to preserve the speaking style of source speech in converted speech. The multi-scale modeling module is designed to model the source speech's style from different levels, i.e., global, local, and frame levels. Besides, an explicit constraint module and simulation training strategy are proposed to directly guide the training of MSM-VC towards the aim of the VC task, i.e., preserving the speaking style of source speech while maintaining the target speaker's timbre. In this section, more details of MSM-VC and its limitations will be discussed as follows.

\subsubsection{Visualization of reconstruction and simulation modes}
To explore the behavior of the simulation mode, the finetune stage's loss curves are visualized in different aspects, including mel reconstruction~($L_{recons}$), style matching~($L_{style}$), and speaker classification~($L_{speaker}$). As shown in Fig.~\ref{figloss}, the two training modes show different behaviors. Specifically, from the beginning of the finetune stage, in terms of $L_{style}$ and $L_{speaker}$, the loss value of the simulation mode is significantly higher than that of the reconstruction mode. This high style matching loss and speaker classification loss obviously drop with the application of the simulation mode. Besides, in terms of the reconstruction loss, the simulation mode does not bring negative effects to the reconstruction process, indicating the effectiveness of the simulation mode in alleviating the mismatch between training and inference and improving the disentanglement ability.

\begin{table}[ht]
\centering
\setlength{\tabcolsep}{1.3mm}
\renewcommand\arraystretch{1.5}
\caption{Model size and inference speed of comparison, ablation, and proposed systems.}
\begin{tabular}{l|c|c}
\hline
 &Trainable Parameters (M) & Real-time Factor  \\ \hline
GST-VC & 4.44 & 0.087 \\
REF-VC & 3.92 & 0.083 \\
Hybrid-VC & 4.24 & 0.088 \\ \hline
MSM-VC & 4.14 & 0.088 \\
$\qquad$w/o Global & 3.97 & 0.087\\
$\qquad$w/o Local & 3.85 & 0.085 \\
$\qquad$w/o Frame & 4.14 & 0.088 \\
$\qquad$w/o Speaker Classifier & 3.75 & 0.088 \\
$\qquad$w/o SER & 4.14 & 0.088 \\
$\qquad$w/o Simulation & 4.14 & 0.088 \\ \hline
\end{tabular}
\label{tab:model_size}
\end{table}

\begin{table}[ht]
\centering
\caption{Comparison of the proposed method with backbone models. 3.17M, 4.44M, and 4.14M are the trainable parameters amount of models.}
\label{tab:basemodel}
\setlength{\tabcolsep}{1.3mm}
\renewcommand\arraystretch{1.5}
\centering
\begin{tabular}{c|c|c|c|c}
\hline
                   \multirow{2}{*}{}    & \multirow{2}{*}{\makecell[c]{Cosine\\Similarity\\$(\uparrow, 0.881)$}}    & \multicolumn{2}{c|}{ \makecell[c]{Pearson\\ Coefficient~$(\uparrow)$}}  &  \multirow{2}{*}{\makecell[c]{Speech\\Quality$(\uparrow)$}} \\ \cline{3-4}
                &  & \multicolumn{1}{c|}{Lf0}      & Energy   &                                                   \\ \hline                   
\makecell[c]{Base-VC (3.17M)} & 0.830  &0.596  & 0.788          &  3.57$\pm$0.091       \\ 
\makecell[c]{Base-VC (conformer 1$\rightarrow$2, \\4.44M)}                & 0.833              &0.603 &0.776    & 3.58$\pm$0.109        \\ 
\makecell[c]{MSM-VC (4.14M)}                 & 0.823               &0.757 &0.968   &3.54$\pm$0.083         \\\hline
\end{tabular}
\vspace{-5pt}
\end{table}

\subsubsection{Model size investigation of different systems}

To investigate the impact of model size on VC performance, the trainable parameters amount among comparison, ablation, and proposed models is shown in Table.~\ref{tab:model_size}. The number of MSM-VC's trainable parameters has no significant increase compared to other comparison systems. Compared with ablation systems, the differences in the trainable parameter amount come from the multi-scale style extraction of the source speech and the generation of constraints for the training process, which bring the enhancement of style modeling and target speaker timber maintenance. Besides, we also investigated the performance of the base model Base-VC, which is a variant of the proposed model by dropping out the proposed multi-scale style extraction and generation constraints, with simply increased parameters. To be specific, we double the layers of the conformer block in Base-VC, referred to as \textbf{Base-VC (conformer 1$\rightarrow$2)}. As shown in Table.~\ref{tab:basemodel}, compared with Base-VC, Base-VC (conformer 1$\rightarrow$2) gets performance gain in speaker similarity and speech quality but worse style similarity. It indicates that simply increasing the model size cannot lead to obvious style modeling performance gain. In MSM-VC, the role of the increased part of trainable parameters is designed to provide more style-related information of source speech from different scales instead of only model capacity. Simply expanding the number of model parameters is unlikely to provide additional style information.

\subsubsection{Limitations}
While experiments have demonstrated the good performance of the proposed model on source style modeling in most scenes, we have to point out that some limitations still exist. Specifically, in some extreme cases, such as speech with high emotional intensity, crying, laughing, shouting, and murmuring, the intelligibility and quality of the converted speech will be greatly affected. The high-quality data of this kind of speech is very difficult to collect, and it is hard to cover all categories simultaneously. Improving the stability and generalization of semantic representation is an interesting topic that needs to be paid more attention to in practice. Furthermore, MSM-VC requires several pre-trained models, e.g., ASR, SER, and vq-wav2vec, which would be disadvantageous in a practical scenario. As presented in Table \ref{tab:model_size}, we tested our model's real-time performance on a single NVIDIA RTX 2080 GPU, achieving a rate of 0.088, which is slightly slower than the comparison methods. While our model is capable of transferring source style to target speakers, it is still non-real-time, limiting its potential applications. Developing a streaming model to accomplish this task would be valuable.

\section{conclusion}
\label{sc:conclusion}
This paper proposes a multi-style modeling approach for the voice conversion task based on a recognition-synthesis framework, which can convey not only the linguistic content but also the speaking style of source speech to the converted speech. In order to obtain comprehensive speaker-irrelevant style representations, the multi-level style modeling module obtains frame-level, local-level, and global-level styles from specific representations. Besides, to directly guide the source style modeling and target speaker timbre preservation of the proposed model, an explicit constraint module consisting of a speaker classifier and a pre-trained speech emotion model is introduced. This explicit constraint module also makes it possible to simulate the style transfer inference process during the training to encourage the speaker and style disentanglement and prevent the mismatch between training and inference. Experimental results demonstrate that the proposed approach achieves superior performance in conveying source speech style while maintaining target speaker timbre and good speech quality.

\bibliographystyle{IEEEtran}
\bibliography{ref.bib}

\vfill

\end{document}